\documentclass{article}

\usepackage[final, nonatbib]{neurips_2021_ml4ps}

\usepackage[utf8]{inputenc} 
\usepackage[T1]{fontenc}    
\usepackage{hyperref}       
\usepackage{url}            
\usepackage{booktabs}       
\usepackage{amsfonts}       
\usepackage{nicefrac}       
\usepackage{microtype}      
\usepackage{xcolor}         

\usepackage{caption}
\usepackage{subcaption}
\usepackage{amsmath}
\usepackage{comment}
\usepackage{listings}
\usepackage{graphicx}
\usepackage{algorithm}
\usepackage{algorithmic}
\usepackage{amssymb}
\usepackage{listings}
\usepackage{bm, bbm}

\title{Implicit Quantile Neural Networks for Jet Simulation and Correction}

\author{%
  Braden Kronheim\\
  Department of Physics\\
  University of Maryland\\
  College Park, MD 20742\\
  \texttt{bkronhei@umd.edu}\\
  \And
  Michelle P. Kuchera\\
  Department of Physics\\
  Davidson College\\
  Davidson, NC 28035\\
  \texttt{mikuchera@davidson.edu}\\
  \And
  Harrison B. Prosper\\
  Department of Physics\\
  Florida State University\\
  Tallahassee, FL 32306\\
  \texttt{hprosper@fsu.edu}\\
  \And
  Raghuram Ramanujan\\
  Department of Mathematics \& Computer Science\\
  Davidson College\\
  Davidson, NC 28035\\
  \texttt{raramanujan@davidson.edu}
}

\begin{document}

\maketitle

\begin{abstract}
  Reliable modeling of conditional densities is important for quantitative scientific fields such as particle physics. In domains outside physics, implicit quantile neural networks (IQN) have been shown to provide accurate models of conditional densities. We present a successful application of IQNs to jet  simulation and correction using the tools and simulated data from the
  Compact Muon Solenoid (CMS) Open Data portal. 
\end{abstract}

\section{Introduction}

The ability to model and sample from multi-dimensional conditional densities is of critical importance in particle physics where almost all phenomena are subject to random fluctuations. One example is the interaction of a jet, a collimated collection of particles (see, for example, \cite{osti_1263408}), with a particle detector. The manner in which a jet interacts with a detector is usually modeled using a Monte Carlo method based on the widely used \texttt{GEANT4} toolkit~\cite{Ivanchenko:2003xp}. \texttt{GEANT4} provides a high-fidelity stochastic simulation of particle interactions with matter, but comes with a high computational cost. In this paper, we demonstrate how given a large sample of jets simulated with \texttt{GEANT4}, the jet response function --- which maps jets from the particle generation level (i.e., before the particles enter the particle detectors) to the observed jets (i.e., jets recorded in the detectors) --- can be accurately modeled with an Implicit Quantile Neural Network (IQN). Moreover, the IQN performs this mapping at a significantly faster rate than the original \texttt{GEANT4}-based simulation. In addition, an IQN can model the inverse problem: the distribution of particle generation-level jets  that could have yielded a given observed jet. The  efficacy of IQNs in modeling multi-dimensional conditional densities is illustrated using simulated jet data provided by the CMS Collaboration at the CERN Open Data Portal\footnote{\url{https://opendata.cern.ch/}}.

\section{Implicit Quantile Neural Networks}
\label{sec:IQN}
Given a training dataset comprising inputs $\bm{x}$ and a target $y$, the \emph{quantile regression} problem is to construct an estimator $f(\bm{x}; \bm{\theta})$ with parameters $\bm{\theta}$ of the quantile function, which is the inverse of the cumulative distribution function $F(y)$, conditioned on $\bm{x}$. We denote a specific quantile by $\tau = F(y)$.

This problem can be cast as an optimization problem \cite{koenker_bassett_1978} in which the average loss,
\begin{equation}
\mathcal{L}(\bm{\theta}) = \sum_{t:y_t \geq f(\bm{x}_t; \bm{\theta})} \tau |y_t - f(\bm{x}_t; \bm{\theta})| + \sum_{t:y_t < f(\bm{x}_t; \bm{\theta})} (1 - \tau) |y_t - f(\bm{x}_t; \bm{\theta})|,
\label{eq:loss}
\end{equation}
over a set of training examples $(\bm{x}_t, y_t)$ is minimized. 

The use of neural networks as the estimator, $f(\bm{x}; \bm{\theta})$, in Eq.~\ref{eq:loss} was first studied by 
White \cite{White_1992} and Taylor \cite{taylor_2000}. More recently, deep neural networks have been used to model the full quantile function by extending the model to include the quantile $\tau$ as an input  \cite{Ostrovski2018AutoregressiveQN, NEURIPS2019_73c03186}. Additionally, Ostrovski \textit{et\,al.~}\cite{Ostrovski2018AutoregressiveQN} introduced a method to handle multi-dimensional probability densities. The method models a multi-dimensional density $p(y^{(1)}, y^{(2)}, \cdots, y^{(n)}|\bm{x})$ as follows
\begin{align}
p(y^{(1)},y^{(2)},\ldots,y^{(n)}|\bm{x}) &= p(y^{(1)} | \bm{x}) \prod_{i=2}^{n} p(y^{(i)} | \bm{x}, y^{(1)}, \ldots, y^{(i-1)}),
\label{eq:pyx}
\end{align}
where  $p(y^{(1)}|\bm{x}), p(y^{(2)}|\bm{x}, y^{(1)}), p(y^{(3)}|\bm{x}, y^{(1)}, y^{(2)}), \ldots, p(y^{(n)}|\bm{x},y^{(1)},y^{(2)},\ldots,y^{(n-1)})$ are 1-dimensional conditional densities.

In order to apply Ostrovski \emph{et al.}'s method to the jet quantile regression problem,  a single training example $(p_T, \eta, \phi, m)\rightarrow(p_T', \eta', \phi', m')$  is unrolled into the four training examples
\begin{align}
(p_T, \eta, \phi, m,1,0,0,0,0,0,0) &\rightarrow(p_T'),\nonumber\\
(p_T, \eta, \phi, m,0,1,0,0,p_T',0,0) &\rightarrow(\eta'),\nonumber\\
(p_T, \eta, \phi, m,0,0,1,0,p_T',\eta',0)&\rightarrow(\phi'),\nonumber\\
(p_T, \eta, \phi, m,0,0,0,1,p_T',\eta',\phi')&\rightarrow(m') ,
\label{eq:examples}
\end{align}
where $p_T, \eta, \phi, m$ are the jet transverse momentum, pseudo-rapidity, azimuthal angle, and mass, respectively. The
one-hot encoding after $p_T, \eta, \phi, m$ in Eq.\,(\ref{eq:examples}) specifies which target is associated with the given unrolled example. To accommodate the form of the data in Eq.\,(\ref{eq:examples}), we use a model $f(\bm{x}, \bm{z}, \bm{y}^\prime, \tau; \bm{\theta})$  that includes inputs $\bm{z}$ and $\bm{y}'$ for the one-hot encoding and the partially specified target vector, respectively.
The quantile $\tau$ is also an input, albeit one that is generated on-the-fly at training time for each example at each epoch, by drawing an independent sample from $U(0, 1)$.

We train a dense, feed-forward network on batches of unrolled examples (augmented with an independent $\tau$ sample) sampled from the training set. At inference time, the trained model is used in an autoregressive manner: the unrolled examples, each with the desired quantile, are provided in the order shown in Eq.\,(\ref{eq:examples}) with the quantities $p^\prime_T, \eta^\prime, \phi^\prime$ now the values \emph{predicted} by the trained model, rather than the values from the test data.

  Since the trained model
 approximates four quantile functions --- one quantile function at a time depending on which one-hot encoding is used --- the model is an implicit approximation of the multi-dimensional conditional density
 $p(\bm{y} | \bm{x})$. The 1-dimensional conditional   densities from which $p(\bm{y} | \bm{x})$ is formed using Eq.\,(\ref{eq:pyx}) can be computed from $p(y^{n} | \bm{x}, y^{(1)}, \cdots, y^{(n-1)}) = (\partial f_n / \partial \tau)^{-1}$, where $f_n$ is the model supplied with the $n^\text{th}$ one-hot encoding.

One problem that often arises in quantile regression is \emph{quantile crossing}, where the approximation to the quantile function is not monotonic. Prior work has attempted to mitigate this problem by imposing constraints on the model architecture or by optimizing novel loss functions\,\cite{cannon_2018, moon2021}. Tagasovska and Lopez-Paz, however, observed that the problem becomes significantly less pronounced when the full quantile function is approximated\,\cite{NEURIPS2019_73c03186}. In this paper, we propose a regularized average loss 
$$ \mathcal{L}' = \mathcal{L} + \lambda \cdot \mathbbm{1}[-f'] (f')^2$$
that further alleviates this problem, 
where $f' = \partial{f}/\partial{\tau}$ and $\lambda$ is a hyperparameter. By penalizing negative gradients of $f$ with respect to $\tau$, the regularization term favors solutions that are monotonically non-decreasing. This is equivalent to the condition $ (f^\prime)^{-1} = p(y | \bm{x}) > 0$.

\section{Applications}
\label{sec:app}
\subsection{Jets}
In this study, IQNs are used to create a fast simulation of jets using the publicly available tools and simulated data provided by the CMS Collaboration at the CERN Open Data Portal. Simulation of particle collisions (typically, proton-proton collisions) at the Large Hadron Collider (LHC)\footnote{\url{https://home.cern/science/accelerators/large-hadron-collider}} begin with the simulation of collisions between proton constituents --- quarks and gluons --- which are collectively referred to as partons.
Partons are not directly observed; therefore, their transformation into observable particles called hadrons (via a process called hadronization) must also be simulated. A particle clustering algorithm, called anti-$k_\text{T}$~\cite{Cacciari_2008}, is then used to cluster the hadrons into jets. In a simulation, the clustering can be performed either on the particles that would be identified from measured tracks and energy deposits in particle detectors, or on the particles before they interact with the detectors. The jets created using the latter method are generally referred to as particle \emph{generator} (or gen-) jets, while those created using the  former method are called \emph{reconstructed} (or reco-) jets. In the examples studied here, gen-jets will be considered the ground truth, i.e., the jets that would be observed with noise-free detectors. The reco-jets, 
are usually subjected to corrections that render their characteristics as similar as possible to those of the gen-jets, on average, using techniques based on experimental and simulated data~\cite{osti_1263408}. The reco-jets without these corrections will be called the \emph{raw} reco-jets, while those with the corrections will simply be called reco-jets.

The simulated CMS jet data \cite{CERN_data1}, as well as the analysis code (which was modified for this study), come from the CERN Open Data Portal \cite{CERN_data2} with Creative Commons CC0 waivers and GPL3 licenses respectively. Approximately 3.9 million simulated jets were extracted from the portal of which $\sim1.3$~million were set aside as test data. Of the remaining $\sim2.6$~million examples, $\sim1.8$~million examples were used as training data, leaving the remaining $\sim200,000$~examples for model validation and hyperparameter tuning. The complete code to reproduce the results in this paper can be found at the linked Github repository\footnote{\url{https://github.com/alpha-davidson/IQNs-for-Jets}} with a GPL3 license.

\begin{figure}
\begin{subfigure}{.5\textwidth}
  \centering
\includegraphics[width=\linewidth]{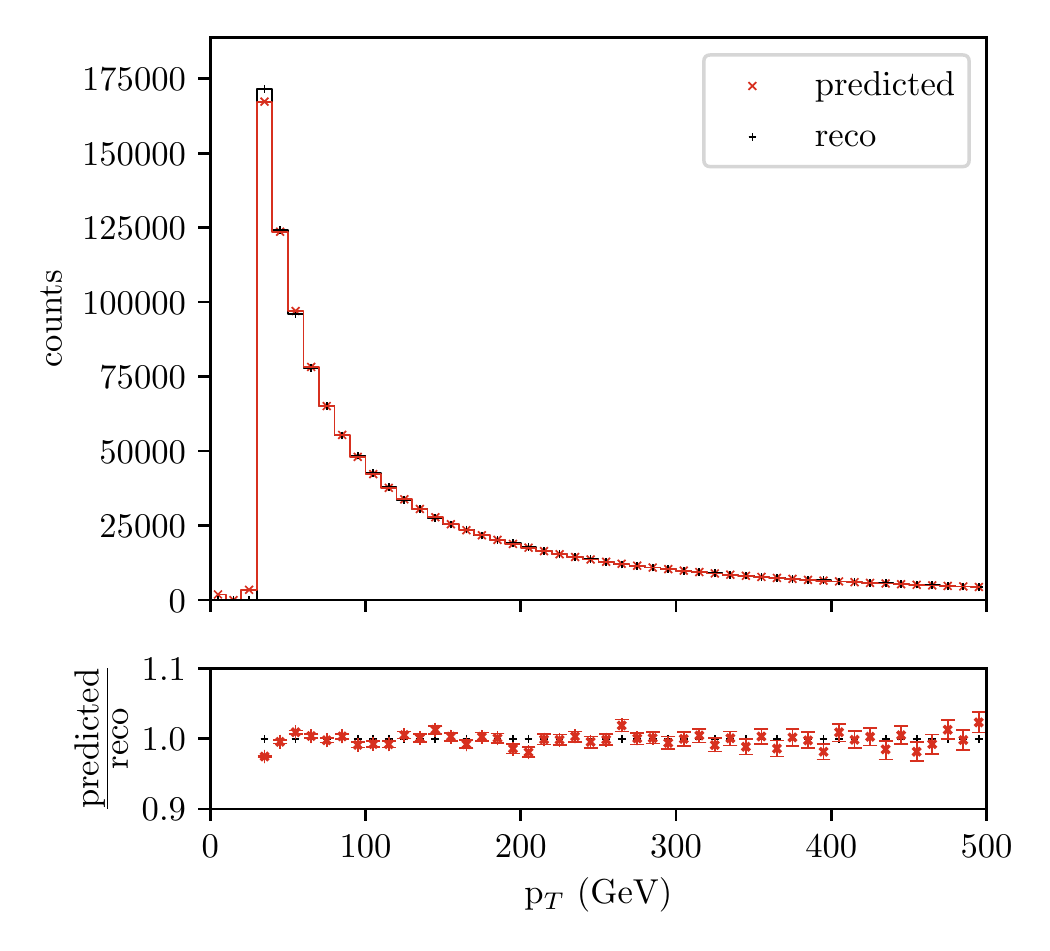}
\end{subfigure}%
\begin{subfigure}{.5\textwidth}
  \centering
  \includegraphics[width=\linewidth]{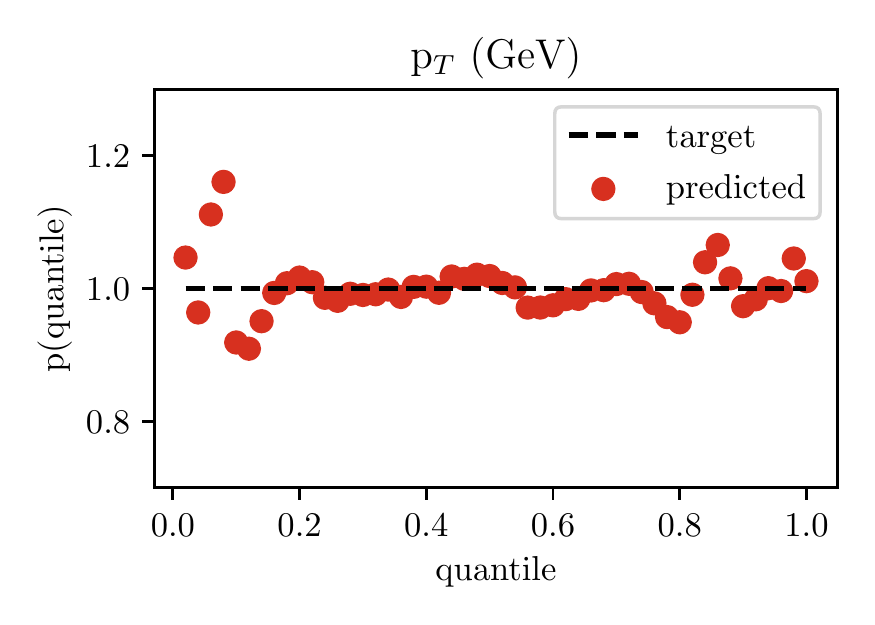}
  \end{subfigure}
\centering
\caption{
\color{black}
(Left) The predicted marginal raw reco-jet $p_T$ spectrum 
is compared with the
 raw reco-jet $p_T$ data from the test set. The predicted histogram displays the median of 1000 predicted marginal distributions.
 The ratio of the displayed spectra is shown in the lower plot. The upper and lower uncertainties are the 0.84 and 0.16 quantiles of the bin count distributions. (Right) The distribution of quantiles, each computed from the true reco-jet transverse momenta using the trained model.}
\label{fig:js}
\end{figure}

\subsection{Detector-Level Jet Simulation}
\label{sec:dljs}

Jet simulation entails mapping particle-level jets (gen-jets) to the
corrected jets (reco-jets) that are observed in particle detectors. The IQN used for this mapping problem is a 5-layer, fully-connected neural network with 50 nodes per layer. We use the Leaky ReLU activation function and train the network using AMSGrad \cite{amsgrad} using only CPU support. A typical training run lasted about 24 hours.

We assess the effectiveness of the trained model by comparing the predicted marginal density of each component of the reco-jet four-momentum ($p^\prime_T, \eta^\prime, \phi^\prime, m^\prime$) with the corresponding true reco-jet marginal density from the data, both integrated over the gen-jet quantities. Specifically, we use the trained model to generate 1000 reco-jet four-momenta predictions for each of the $10^6$ gen-jets from the test set. From these $10^9$ predictions conditioned on the gen-jet inputs, we derive 1000 marginal distributions for each jet parameter, by aggregating over randomly sampled batches of size $10^6$. We present the median of these 1000 marginal distributions for $p_T$ in the left panel of Fig.~\ref{fig:js}, alongside the marginal distribution computed from the test data. We see that the predicted marginal distributions are nearly modeled within uncertainty. We obtained similar results for the other three jet variables as well, but have omitted them in the interest of space. 

Additionally, we perform the following closure test to validate our approach. We construct a cumulative distribution function (cdf)  $F(y^\prime)$ for each of the $10^6$ gen-jets in the test set, where $y' \in \{p^\prime_T, \eta^\prime, \phi^\prime, m^\prime\}$, for each component of the reco-jet momentum vector from our model's $10^9$ predictions. We then use this induced cdf to map each reco-jet measurement from the test set to a quantile. If the IQN's modeling of the quantile function, and therefore the conditional distribution, is accurate, then we should expect the distribution of these quantiles to follow $U(0, 1)$. We see that this is indeed the case, as shown in the right panel of Fig.~\ref{fig:js}. Once again, we have chosen to only present the results for $p_T$ and omit the qualitatively similar results obtained on the other three variables.

\begin{figure}
\begin{subfigure}{.5\textwidth}
  \centering
\includegraphics[width=\linewidth]{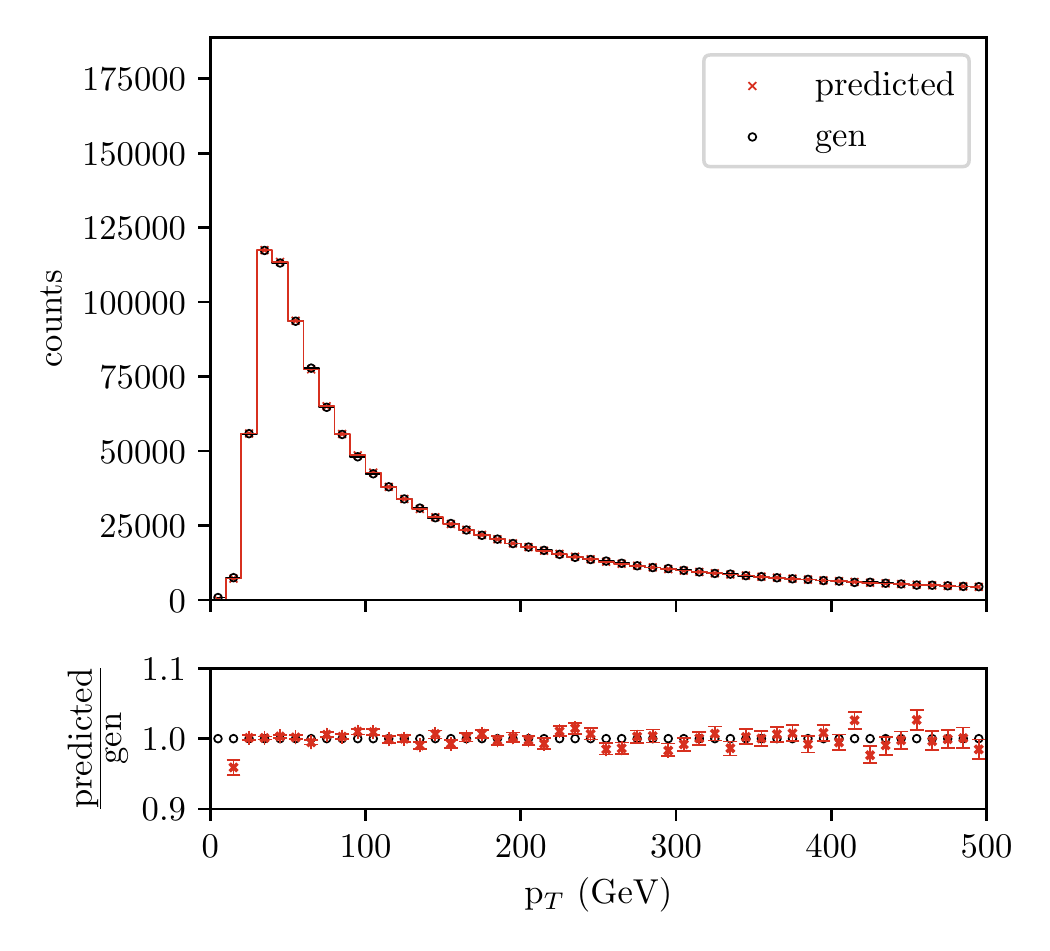}
\end{subfigure}%
\begin{subfigure}{.5\textwidth}
  \centering
  \includegraphics[width=\linewidth]{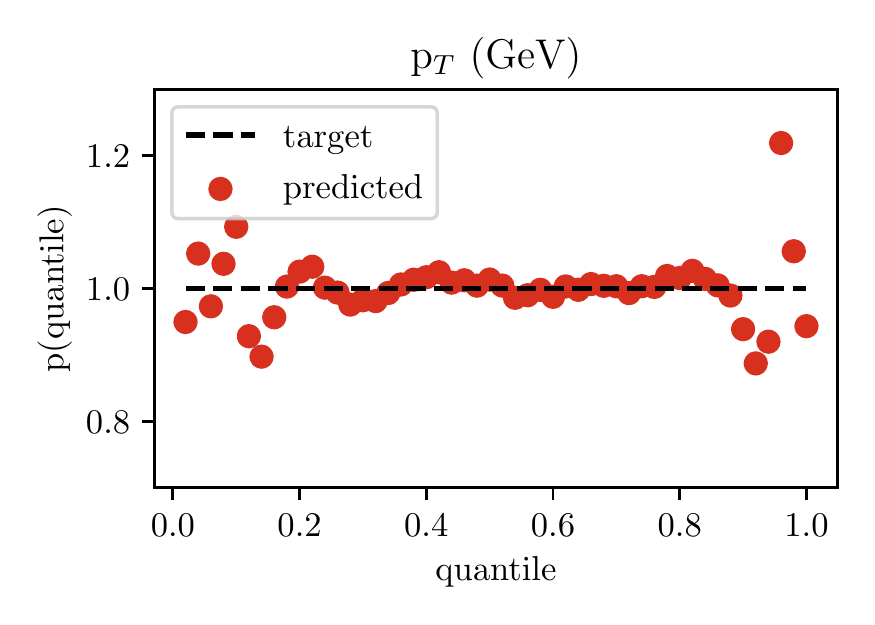}
  \end{subfigure}
\centering
\caption{ (Left) The predicted gen-jet $p_T$ spectrum compared with the
true gen-jet $p_T$ spectrum. Both spectra are integrated over the raw reco-jet quantities. The ratio of the spectra is shown in the lower plot. The  uncertainties reflect the statistical fluctuations in the 1000 predicted spectra.
(Right) The distribution of quantiles, each computed from the gen-jet transverse momenta in the test set using the trained model.
 \color{black}}
\label{fig:jc}
\end{figure}

\subsection{Jet Inverse Problem}
An IQN can also be used to
to solve the jet inverse problem:  what gen-jets are consistent with being the progenitors of a given raw reco-jet? Such a function could be useful, for example, in estimating the particle-level spectrum of a jet quantity from the corresponding observed  spectrum. In Fig.\,\ref{fig:jc}, we show the marginal densities for the predicted gen-jet $p_T$ spectra, the test gen-jet data, and their ratio, computed using the protocol outlined in Sec.~\ref{sec:dljs}, i.e. using $10^9$ predictions divided into 1000 batches of $10^6$ each. The figure also shows the distribution of quantiles, this time computed from the true gen-jet $p_T$. The predicted spectra are again nearly within uncertainty across $p_T$, and a closure test is performed with similar results as the forward predictions. 

\section{Conclusions}
\label{sec:conclusion}

In this work, we presented an application of IQNs to the tasks of jet correction and simulation. Our IQN architecture comprises a feed-forward, fully-connected neural network, which is straightforward to train, particularly when compared to other generative modeling techniques such as GANs that tend to be more unstable and require more careful tuning. Our approach utilizes a novel regularization term that helps mitigate quantile crossing, without interfering with the convergence of the network. The trained IQNs approximate the marginal densities of the jet variables nearly within the uncertainty of the predictions, across the parameter space. Confirming that the multi-dimensional conditional densities $p(\bm{y} | \bm{x})$ are modeled correctly beyond the correspondence shown in the closure tests presents additional challenges, as those distributions are not explicitly present in the data, and is the focus of ongoing study.

\section{Impact Statement}
Conditional densities are ubiquitous in particle physics. For example, they appear in statistical models $p(\bm{d} | \bm{\mu}, \bm{\nu})$, where $\bm{d}$ are observable data and $\bm{\mu}$ and $\bm{\nu}$ are parameters of interest and nuisance parameters, respectively. They also appear in response functions $r(\bm{y} | \bm{x}) $ that appear in multi-dimensional integrals of the form $o(\bm{y}) = \int r(\bm{y} | \bm{x}) \, u(\bm{x}) \, d\bm{x}$ that
map an unobserved spectrum $u(\bm{x})$ to an observed spectrum $o(\bm{y})$ of which the jet 4-momenta spectra are a typical example. 
There is a renewed push in particle physics to publish full statistical models. IQNs provide a simple and effective way to both encapsulate statistical models and compute them very quickly, as well as to model the numerous response functions that appear in the analysis of particle physics data at the Large Hadron Collider and other particle physics research facilities. Finally, IQNs could be the basis of very fast detector simulations that are automatically abstracted from existing and future high-fidelity \texttt{GEANT4}-based event simulations.

From a machine learning standpoint, this paper introduces a novel regularization term to enforce the requirement that the neural network learn a monotonically increasing quantile function. This modification to the standard quantile regression loss function is potentially useful to any application that requires a model to produce a statistically meaningful quantile function and cumulative distribution function post-training.

\begin{ack}
This work was supported by the National Science Foundation under Cooperative Agreement OAC-1836650.

\end{ack}

\bibliographystyle{unsrt}
\bibliography{main.bib}

\end{document}